\documentclass[11pt]{article}
\usepackage{times}
\usepackage{geometry}
\usepackage[utf8]{inputenc}
\usepackage{enumitem,amssymb}
\usepackage{ragged2e}
\newlist{thematic}{itemize}{8}
\setlist[thematic]{label=$\square$}
\usepackage{pifont}
\usepackage{fancyhdr}
\usepackage[USenglish]{babel}
\usepackage[utf8]{inputenc}
\usepackage[T1]{fontenc}
\usepackage{epsfig}
\usepackage{wrapfig}
\usepackage{color}
\usepackage{tabularx}

\usepackage[vflt]{floatflt}
\usepackage{longtable}
\usepackage{caption}
\usepackage{jheppub}   
\usepackage[dvipsnames,svgnames,x11names]{xcolor}
\usepackage[all]{hypcap} 
\usepackage{amssymb,amsmath}
\usepackage{longtable}
\usepackage{amssymb}
\usepackage{amsmath, xspace}
\usepackage{graphics,graphicx} 
\usepackage{rotating}
\usepackage{color}
\usepackage{siunitx}
\usepackage{xcolor}
\usepackage{multirow}
\usepackage{subfigure}
\usepackage{sectsty}
\usepackage[outercaption]{sidecap}
\sidecaptionvpos{figure}{c}

\usepackage{enumitem}
\setlist[enumerate]{itemsep=0pt, parsep=0pt}
\setlist[itemize]{itemsep=0pt, parsep=0pt}

\definecolor{DarkGreen}{rgb}{0.0, 0.3, 0.0}
\definecolor{purple}{rgb}{0.5, 0.0, 0.5}
\definecolor{red}{rgb}{1, 0.0, 0.0}
\definecolor{green}{rgb}{0, 1.0, 0.0}

\def\3he{$^3{\rm He}$}

\def\lsim{\mathrel{\lower2.5pt\vbox{\lineskip=0pt\baselineskip=0pt
           \hbox{$<$}\hbox{$\sim$}}}}

\def\gsim{\mathrel{\lower2.5pt\vbox{\lineskip=0pt\baselineskip=0pt
           \hbox{$>$}\hbox{$\sim$}}}}

\begin{document}
\raggedright
\huge
Revealing the baryon cycle in Galaxy Clusters: connecting galaxy dynamics and gas thermodynamics using (sub-)mm-wave and optical IFU surveys 
\linebreak
\bigskip
\normalsize

\textbf{Authors:} 
Francisco M. Montenegro-Montes$^{1\star}$,
Patricia Sánchez-Blázquez$^1$, Tony Mroczkowski$^2$, Armando Gil de Paz$^1$, Cristina Catalán-Torrecilla$^1$, Marie-Lou Gendron-Marsolais$^3$, Paula Macías-Pardo$^1$, Beatriz Callejas-Córdoba$^1$, Alfredo Montaña$^4$, Juan F. Macías-Pérez$^5$ and Susana Planelles$^6$.
\linebreak
\linebreak
$^1$ Institute of Particle and Cosmos Physics, Univ. Complutense de Madrid (IPARCOS-UCM), Spain\\
$^2$ Institute of Space Sciences, Consejo Sup. de Investigaciones Científicas (ICE-CSIC), Spain\\
$^3$ Département de physique, de génie physique et d’optique, Université Laval, Québec (QC), Canada\\
$^4$ Instituto Nacional de Astrofísica Óptica y Electrónica (INAOE), Mexico\\
$^5$ Université Grenoble Alpes, CNRS, GrenobleINP, LPSC-IN2P3, France\\
$^6$ Departament d’Astronomia i Astrofísica, Universitat de València, Spain\\
$^{\star}$ E-mail: fmontene@ucm.es
\linebreak
\linebreak

\textbf{Science Keywords:} 
cosmology: circumgalactic medium; galaxies: clusters; cosmology: large-scale structure of universe
\linebreak

 \captionsetup{labelformat=empty}
\begin{figure}[h]
   \centering
\includegraphics[width=.9\textwidth]{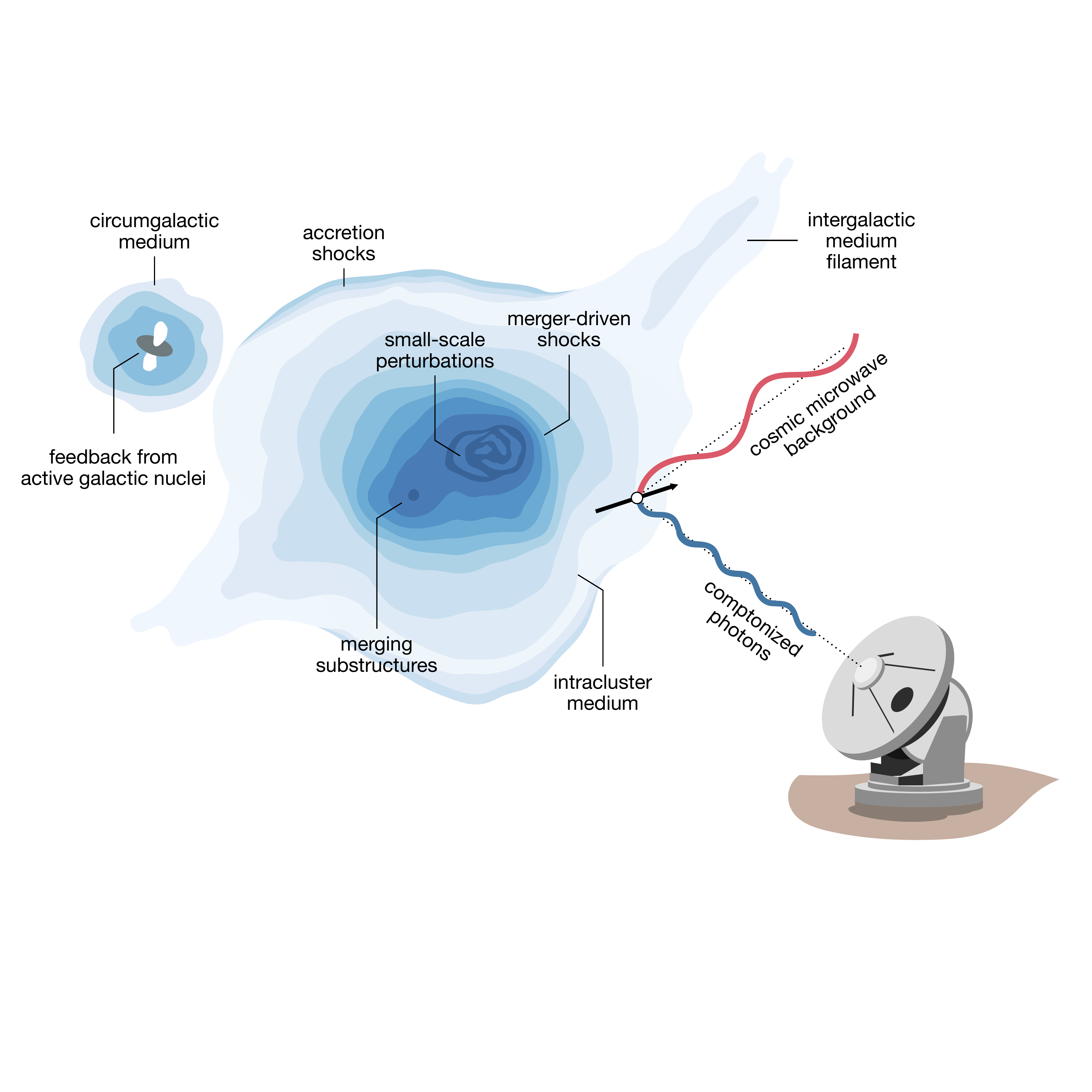}
\vspace{-30mm}
   \caption{Cartoon depicting some of the rich cosmic structures accessible through high-resolution mm-wave observations of the Sunyaev-Zeldovich (SZ) effect.  Image from Di~Mascolo et al.\ 2025 [2].}
\end{figure}
\vspace{-15mm}

\setcounter{figure}{0}
\captionsetup{labelformat=default}


\pagebreak

\section*{Abstract}
\vspace{-3mm}

Observations in the visible and near infrared are transforming our view of the processes affecting galaxy evolution, much of which is dominated by interactions with the large scale environment. Yet a complete picture is missing, as no corresponding high resolution view of the warm/hot intracluster, circumgalactic, and intergalactic media exists over large areas and a comparably broad range of redshifts. Combined with wide-field optical IFU surveys such as CATARSIS, a large diameter sub-mm telescope with a degree-scale field of view would enable a joint view of galaxy dynamics and gas thermodynamics, transforming our understanding of environmental processes.
\vspace{-3mm}

\section{Scientific context and motivation}
\vspace{-3mm}

Galaxy clusters are the end points of hierarchical structure formation, containing the majority of baryons in the form of hot, diffuse plasma. The thermodynamic and dynamical state of this intracluster medium (ICM) encodes the history of gravitational collapse, feedback, and accretion. Over the last two decades, X-ray and low-resolution (arcminute) Sunyaev-Zeldovich (SZ) surveys have provided azimuthally averaged density and pressure profiles of the ICM (e.g. [1,6,7]). These measurements lack the angular resolution and mapping speed needed to characterize the spatially resolved structure of clusters, particularly their outskirts and filaments where most baryonic mass is currently being accreted.\\
\vspace{2.5mm}
Crucially, X-ray observations face fundamental limitations in the low-density regions where cluster growth occurs. X-ray surface brightness scales as $n_{e}^{2}$, causing the emission to fall off steeply beyond $r_{500}$, and temperature measurements become photon-starved in the outskirts [10]. This leaves present and future X-ray missions (e.g. NewAthena, expected to launch by the late 2030s) unable to probe in detail these external regions (see e.g. [4]).\\
\vspace{2.5mm}
At the same time, surveys in the visible using integral field units (IFUs), such as CATARSIS, planned for the next decade with TARSIS [3], will revolutionize our understanding of galaxy dynamics and environmental transformation within clusters. CATARSIS will provide unbiased redshift and kinematic information for thousands of galaxies per system, up to five times the virial radius [8]. These data trace the dark matter assembly and galaxy infall history independent of hydrostatic assumptions.\\
\vspace{2.5mm}
What remains missing, therefore, is the high-resolution thermodynamic context of this dynamical and multi-component picture: How do the pressure, entropy, and turbulence of the ICM evolve across the large spatial scales mapped by CATARSIS? How are the galaxies’ orbits and star-formation histories linked to the hot gas in clusters and the filaments around them? 
Truly addressing these questions requires high-resolution SZ and deep continuum imaging that extends across entire clusters and their infall regions. 
Achieving this demands a large-aperture ($\geq$50 m) single-dish (sub)millimetre telescope, with a degree-scale field of view and broad frequency coverage from the millimetre to sub-millimetre regime, capable of mapping the ICM over all relevant spatial and physical scales.\\

\vspace{-3mm}
\section{Science case}
\vspace{-2mm}

Understanding cluster assembly and the baryon cycle requires answering a set of tightly connected questions that link galaxy dynamics to the thermodynamics of the surrounding gas:
\begin{enumerate}
    \item {\bf How and where do clusters accrete baryons from the cosmic web?} What are the physical boundaries--the splashback radius and the accretion shock--and how do they evolve with cluster mass and accretion rate?
    \item {\bf What drives the transformation of infalling galaxies?} How do the local ICM pressure and turbulence regulate ram-pressure stripping, strangulation, and the quenching of star formation?
    \item {\bf Where are the missing baryons?} Are the outskirts depleted, or do non-thermal pressure and gas clumping hide a large fraction of the hot plasma?
\end{enumerate} 
Addressing these questions demands spatially resolved measurements of pressure, entropy, and gas motions across the same large scales provided by CATARSIS, which in turn places clear demands on the capabilities of a future (sub-)mm facility:

\begin{enumerate}
    \item {\bf Mapping power.} A large $\sim$50-m dish can deliver 5$^{\prime\prime}$–15$^{\prime\prime}$ resolution in the various SZ bands (90-150-350 GHz), ideally over $\ge$ 1$^{\circ}$ fields. This allows efficient measurement of pressure, temperature, and entropy distributions well beyond the virial radius, e.g.\ 2–3 r$_{200}$.
    \item {\bf Direct detection of non-thermal structure.} Multi-frequency SZ observations will separate thermal, kinetic, and relativistic components, revealing turbulence, bulk flows, and shock fronts that generate hydrostatic bias in X-ray–based mass estimates [6].
    \item {\bf Synergy with CATARSIS results.} CATARSIS provides complementary galaxy velocity fields, stellar populations, and star-formation diagnostics out to the same scales. Cross-correlation with resolved SZ maps enables: (i) Measurement of mass-accretion rates by comparing the dynamical splashback radius (from galaxy phase-space) to thermodynamic discontinuities (from pressure maps); (ii) Hydrostatic-bias calibration, through direct comparison of caustic and SZ-derived mass profiles; (iii) Quantification of environmental quenching efficiency as a function of measured ICM pressure rather than projected radius; (iv) Recovery of the dust-obscured star formation in infalling galaxies and stripped material via sub-mm continuum observations, enabling a complete assessment of environmental quenching.
\end{enumerate}
   
\begin{figure}[h]
   \centering
\includegraphics[width=.95\textwidth]{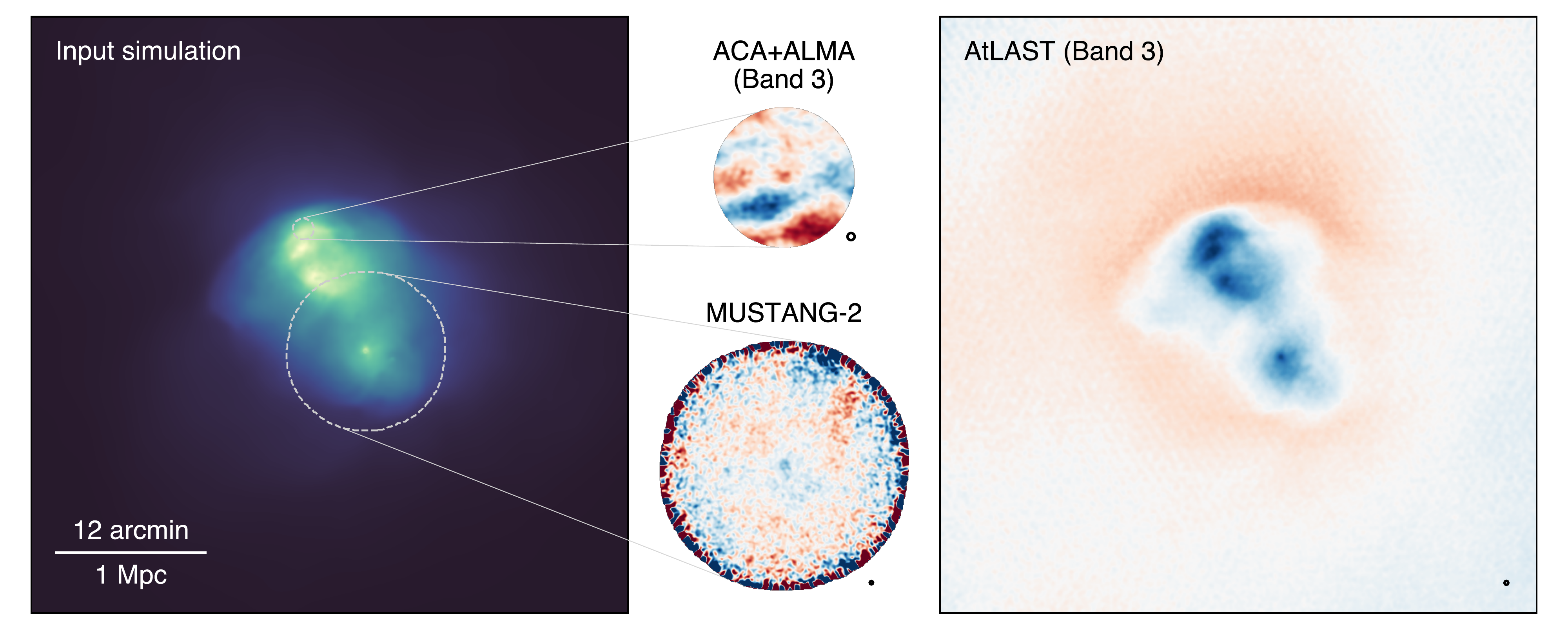}
   \caption{{\bf Figure 1.} The image shows a cluster (left) from the Dianoga cosmological simulation suite, along with 90~GHz mock observations made using ALMA Band 3 (upper middle) and the 100-meter Green Bank Telescope with MUSTANG-2 (lower middle). These illustrate some limitations of current facilities for stuyding the large-scale gas features that trace dynamical processes in galaxy clusters. The right panel presents a mock AtLAST observation at 90 GHz using a modest first-generation instrument; a multi-band configuration will further improve angular resolution and component separation. Image from [2].}
   \label{fig:mockobs}
\end{figure}

No current or planned mm-wave facility (ALMA, CCAT/FYST, MISTRAL, MUSTANG-2, NIKA2, TolTEC, Simons Observatory, or SPT-3G+) combines high mapping speed, wide-field coverage, and multi-band SZ sensitivity at arcsec-resolution on degree scales. ALMA provides exquisite high-resolution imaging but cannot map entire clusters, both due to the small field of view and the filtering effects intrinsic to interferometry; it is worth noting SKAO, ngVLA, and upgrades to ALMA will also fail to solve this. 
Meanwhile, bolometer or kinetic inductance detector arrays like TolTEC and other wide-field cameras achieve similar angular resolution at 90–150 GHz but over much smaller instantaneous fields of view ($\sim 5$ arcminutes), limiting both depth, area covered, and spatial dynamic range. These effects are shown in Figure \ref{fig:mockobs}.

 \captionsetup{labelformat=empty}
\vspace{-3mm}
\section{Technical requirements}

\begin{table}[ht!]
\centering
\caption{\textbf{Table 1} Key Technical requirements to address the CATARSIS-SZ science case.}
\label{tab:technical_requirements}
\renewcommand{\arraystretch}{1.3}
\begin{tabularx}{\textwidth}{>{\raggedright\arraybackslash}p{3.2cm} X}
\hline
\textbf{Parameter} & \textbf{Requirement / Motivation} \\
\hline
\textbf{Frequency coverage} & 
$90$--$350~\mathrm{GHz}$, encompassing the thermal SZ decrement ($\sim90~\mathrm{GHz}$), 
the main SZ bands ($\sim90-150~\mathrm{GHz}$), the null ($\sim 220~\mathrm{GHz}$), and the increment ($\sim 270-350~\mathrm{GHz}$) for component separation. 
Higher frequencies ($\geq350~\mathrm{GHz}$) will aid in characterizing dust emission from galaxies and the ICM. 
For massive systems with $k_B T_e \gtrsim 5$~keV, higher frequencies will aid in precise temperature maps through relativistic corrections to the thermal SZ effect (see e.g. [2]). \\

\textbf{Angular resolution} & 
$\lesssim 10^{\si{\arcsecond}}$ at $150~\mathrm{GHz}$ (typical for a 50-m dish), sufficient to resolve ICM substructures and 
to separate cluster galaxies from diffuse gas emission. This resolution should be maintained over large mapped fields. 
The higher resolutions ($\lesssim 2^{\si{\arcsecond}}$ at $>600~\mathrm{GHz}$) available at higher frequencies will aid in mitigation of cosmic infrared background sources. \\

\textbf{Mapping speed} & 
$>1000~\mathrm{deg}^2~\mathrm{hr}^{-1}~\mathrm{mJy}^{-2}$ (approximately two-three orders of magnitude faster than current single-dish facilities) to reach low surface-brightness (few $\mu$Jy/beam level) sensitivity in the cluster outskirts and connecting filaments. \\

\textbf{Field of view} & 
$\geq1^{\circ}$ instantaneous FoV to recover scales $\gtrsim3\,r_{200}$ for clusters at $z\lesssim0.3$ [9]. \\

\textbf{Dynamic range} & 
Sensitivity to faint SZ signals (a few \si{\micro\kelvin}) in the presence of bright sub-mm sources, requiring simultaneous 
multi-band imaging and robust source subtraction algorithms. \\

\textbf{Data handling} & 
Joint analysis pipelines combining AtLAST SZ and continuum maps with optical IFU data (velocity fields, 
star-formation rates, and stellar populations) on matched astrometric grids. 
Hierarchical Bayesian frameworks will merge thermodynamic and dynamical information to reconstruct 
mass and pressure fields. \\
\hline
\end{tabularx}
\end{table}
The Atacama Large-Aperture Submillimeter Telescope (AtLAST; [5]) is uniquely suited to address this science case. Table \ref{tab:technical_requirements} summarizes the essential technical requirements needed to achieve it.

In summary, AtLAST’s wide-field, multi-band, high-sensitivity sub-mm capabilities offer the indispensable thermodynamic counterpart to the dynamical mapping provided by CATARSIS. Together, these facilities will transform galaxy clusters from static cosmological probes into dynamic laboratories where the physics of baryonic accretion, feedback, and environmental transformation can be directly measured.

\section{References}
\vspace{-4mm}
[1] Arnaud, M., Pratt, G. W. et al.\ (2010). A\&A, 517, A92. \href{https://doi.org/10.1051/0004-6361/200913416}{doi:10.1051/0004-6361/200913416}\\
\vspace{1mm}
[2] Di Mascolo, L., Perrott, Y. et al.\ (2025), ORE, Vol. 4, 113.\ \href{https://doi.org/10.12688/openreseurope.17449.2}{doi:10.12688/openreseurope.17449.2}\\
\vspace{1mm}
[3] Gil de Paz, A. et al.\ (2024), Proc. of the SPIE, Vol 13096, id. 1309620. \href{https://doi.org/10.1117/12.3016123}{doi:10.1117/12.3016123}\\
\vspace{1mm}
[4] Lotti, S., D'Andrea, M. et al.\ (2021), ApJ, 909, 111.\ \href{https://doi.org/10.3847/1538-4357/abd94c}{doi:10.3847/1538-4357/abd94c}\\
\vspace{1mm}
[5] Mroczkowski, T., Gallardo P.A. et al.\ (2025), A\&A, 694, A142.\href{https://doi.org/10.1051/0004-6361/202449786}{doi:10.1051/0004-6361/202449786}\\
\vspace{1mm}
[6] Mroczkowski, T., Nagai et al.\ (2019). Space Science Rev., 215, 17. \href{https://doi.org/10.1007/s11214-019-0581-2 }{doi:0.1007/s11214-019-0581-2}. \\
\vspace{1mm}
[7] Planck collaboration et al.\ (2013). A\&A, 550, A131. \href{https://doi.org/10.1051/0004-6361/201220040}{doi: 10.1051/0004-6361/201220040}\\
\vspace{1mm}
[8] Sánchez-Blázquez, P., Gil de Paz et al.\ (2025), Highlights of Spanish Astrophysics XII, 103. \href{https://www.sea-astronomia.es/sites/default/files/archivos/proceedings16/ORALES/Galaxias/GX_SanchezBlazquez_P.pdf}{ISBN: 978-84-09-70321-0}\\
\vspace{1mm}
[9] van Marrewijk, J., Morris, T. et al.\ (2024), OJAp, 7, 118.\ \href{https://doi.org/10.33232/001c.127571}{doi:10.33232/001c.127571}\\
\vspace{1mm}
[10] Walker, S. et al.\ (2019). Space Science Rev., 215, 7. \href{https://doi.org/10.1007/s11214-018-0572-8}{doi:10.1007/s11214-018-0572-8}\\
\end{document}